\documentstyle[12pt,world_sci]{article}
\font\tenbf=cmbx10
\font\tenrm=cmr10
\font\tenit=cmti10
\font\elevenbf=cmbx10 scaled\magstep 1
\font\elevenrm=cmr10 scaled\magstep 1
\font\elevenit=cmti10 scaled\magstep 1

\renewenvironment{thebibliography}[1]
 { \elevenrm
   \begin{list}{\arabic{enumi}.}
    {\usecounter{enumi} \setlength{\parsep}{0pt}
     \setlength{\itemsep}{3pt} \settowidth{\labelwidth}{#1.}
     \sloppy
    }}{\end{list}}

\begin{document}
\begin{flushright}
{\normalsize Yaroslavl State University\\
             Preprint YARU-HE-94/03\\
             hep-ph/9409321} \\[3cm]
\end{flushright}
\begin{center}{{\tenbf COULD VECTOR LEPTOQUARKS BE RATHER LIGHT?}
\vglue 1.0cm
{\tenrm A.V.~KUZNETSOV and N.V.~MIKHEEV
\footnote[1]
{E-mail addresses: phystheo@univ.yars.free.net, phth@cnit.yaroslavl.su}
\\}
\baselineskip=13pt
{\tenit
Division of Theoretical Physics, Department of Physics,\\}
\baselineskip=13pt
{\tenit
Yaroslavl State University, Sovietskaya 14,\\}
\baselineskip=13pt
{\tenit
Yaroslavl, 150000 Russia\\}
\vglue 0.8cm

{\tenrm ABSTRACT}}
\end{center}
\vglue 0.3cm
{\rightskip=3pc
 \leftskip=3pc
 \tenrm\baselineskip=12pt
 \noindent

A possible Standard Model extension of the Pati-Salam type with a lepton
number as the fourth color is reexamined. A new type of mixing in the
interaction of the $SU(4)_V$--leptoquark with quarks and leptons is shown to
be required. An additional arbitrariness of the mixing parameters could
allow to decrease noticeably the lower bound on the
leptoquark mass $M_X$ originated from the $\pi$ and $K$ decays
and the $\mu e$ conversion.
The only mixing independent bound emerging from the
cosmological limit on the $\pi^0 \rightarrow \nu \bar{\nu}$ decay
width is $M_X > 18~TeV$.

\vglue 3cm}

\begin{center}
{\it Talk given at the VIII International Seminar "Quarks-94",\\ Vladimir,
Russia, May 11-18, 1994}
\end{center}
\newpage

\baselineskip=14pt
\elevenrm

Although the Standard Model predictions are in good agreement with
experiment now, see e.g. ref.~\citenum{A}, a hope for new physics
beyond the Standard Model undoubtedly exists. If the
consecutive restoration of higher symmetries with an energy increase
is assumed, one can speak about some stairway of the symmetries and the
corresponding mass levels. The question is pertinent what the next stair
after the Standard Model could be? Considering the concept widely covered
in the literature of low-energy supersymmetry one can point out that the
symmetry of fermions and bosons would be higher than the symmetry within
the fermion sector, namely, the quark-lepton symmetry. There could also
be expected the symmetry within the boson sector of gauge weak bosons
interacting with the left and right currents, namely, the left-right
symmetry. Thus the supersymmetry restoration could be connected, in our
opinion, with higher
mass scale then the others. We should like to discuss one of the possibilities
when the left-right symmetry restores at an appreciably higher mass scale
then the quark-lepton one. So, we take the minimal symmetry of the Pati-Salam
type with a lepton number as the fourth color~\cite{PSm} based on the gauge
group $SU(4)_V \otimes SU(2)_L \otimes G_R$. The fermions are combined into
the following representations of $SU(4)_V$ subgroup

\begin{equation}
\left ( \begin{array}{c} u^1 \\ u^2 \\ u^3 \\ \nu \end{array}
\right )_i \, , \qquad \left (
\begin{array}{c} d^1 \\ d^2 \\ d^3 \\ \ell \end{array} \right )_i \, ,
\qquad (i=1,2,3) , \label{eq:q}
\end{equation}

\noindent where the $i$ index labels the fermion generations.
Some attractive features of the model should be pointed out:

\begin{enumerate}

\item Let us remember that some quark-lepton symmetry is necessary for
the renormalizability of the Standard Model, namely, the fermions are bound
to be combined into generations for the cancellation of the triangle anomalies.

\item The proton decay is absent in this model.

\item The model gives a natural explanation for the quark fractional
hypercharge. Really, the 15-th generator of $SU(4)$ can be written in the
form:

\begin{equation}
T_{15} \; = \; \sqrt{\frac{3}{8}} \; diag \left ( \frac{1}{3} \, , \,
\frac{1}{3} \, , \, \frac{1}{3} \, , \, -1 \right ) \; = \sqrt{\frac{3}{8}}
\; Y_L.
\label{eq:T}
\end{equation}

It is traceless and the values of the left hypercharge appear to be placed
on the diagonal. Let us call it the vector hypercharge, $Y_L = Y_V$.

\item Let us suppose that $G_R \, = \, U(1)_R$~\cite{S} and try to find
the values of the right hypercharge $Y_R$ for quarks and leptons.
Recall that the values of the
hypercharge of left and right, and $up$ and $down$ quarks and leptons
in the Standard Model are the following

\begin{equation}
Y_{SM} \; = \; \left \{
\begin{array}{c} \left (\begin{array}{c} \frac{1}{3} \\ \\ \frac{1}{3}
\end{array} \right ) \quad for \; q_L ; \\ \\
\left (\begin{array}{c} \frac{4}{3} \\ \\ -\frac{2}{3}
\end{array} \right ) \quad for \; q_R ; \end{array}
\begin{array}{c} \left (\begin{array}{c} - 1 \\ \\ - 1
\end{array} \right ) \quad for \; \ell_L \\ \\
\left (\begin{array}{c} 0 \\ \\ - 2
\end{array} \right ) \quad for \; \ell_R \end{array}
\right \}.
\label{eq:Y}
\end{equation}

If we write now $Y_{SM} \, = \, Y_V \, + \, Y_R$, then the values of the right
hypercharge $Y_R$ occur to be equal $\pm 1$ for the $up$ and $down$ fermions,
both quarks and leptons.
It is tempting to interpret this fact as the evidence for the
right hypercharge to be actually the doubled third component of the right
isospin. Hence the $G_R$ group is possibly $SU(2)_R$ and thus
the symmetry of the $SU(4)_V \otimes SU(2)_L \otimes SU(2)_R$ type could be
the next stair of the above-mentioned stairway.

\end{enumerate}

The most exotic object of the Pati--Salam type symmetry is the charged and
colored gauge $X$
boson named leptoquark. Its mass $M_X$ should be the scale of reducing
of $SU(4)_V$ to $SU(3)_c$. The bounds on the vector leptoquark mass
{}~\cite{PDG} were obtained from the data on the $\pi \rightarrow e \nu$
decay to be $m_X \, > \, 125 \, TeV$~\cite{Sha}, and from the upper limit
on $K^0_L \rightarrow \mu e$
decay to be $m_X \, > \, 350 \, TeV$~\cite{Des}. In fact, these estimations
were not comprehensive because the phenomenon of a mixing in the lepton-quark
currents was not considered there. It can be shown that such a mixing
inevitably occurs in the theory. Really, three fermion generations are
combined into the \{4,2\} representations of the semi-simple group
$SU(4)_V \otimes SU(2)_L$ of the type

\begin{equation}
\left ( \begin{array}{c} u^c \\ \nu \end{array} \,
\begin{array}{c} d^c \\ \ell \end{array}
\right )_i , \qquad (i=1,2,3) , \label{eq:d}
\end{equation}

\noindent where $c$ is the color index to be further omitted. The mixing
in the quark interaction with the $W$ bosons being depicted by
the Cabibbo--Kobayashi--Maskawa matrix is sure to exist in Nature. Therefore,
at least one of the states $u$ or $d$ in~(\ref{eq:d}) is not the mass
eigenstate. It can be easily seen that in the general case, none of the
components in~(\ref{eq:d}) is the mass eigenstate because of arising of
the mixing at the loop level. For example, if we start from the $d$ state to
be diagonal with respect to mass it becomes non-diagonal when the one-loop
transitions of the type $d \rightarrow u + W^- \rightarrow d'$ are taken
into account. It leads to the non-diagonal transitions $\ell \rightarrow
d + X \rightarrow \ell'$ through the quark-leptoquark loop. Consequently,
it is necessary for the renormalizability of the model to include all kinds
of mixing at the tree-level. Due to the identity of the three representations
{}~(\ref{eq:d}) they always could be regrouped so that one of the components
was
diagonalized with respect to mass. The diagonalization of the charged
lepton mass matrix seems to be the most natural, and the
representations~(\ref{eq:d}) can be rewritten in the form

\begin{equation}
\left ( \begin{array}{c} u \\ \nu \end{array} \,
\begin{array}{c} d \\ \ell \end{array}
\right )_\ell \; = \;
\left ( \begin{array}{c} u_e \\ \nu_e \end{array} \,
\begin{array}{c} d_e \\ e \end{array}
\right ) , \;
\left ( \begin{array}{c} u_\mu \\ \nu_\mu \end{array} \;
\begin{array}{c} d_\mu \\ \mu \end{array}
\right ) , \;
\left ( \begin{array}{c} u_\tau \\ \nu_\tau \end{array} \;
\begin{array}{c} d_\tau \\ \tau
\end{array} \right ) , \label{eq:d2}
\end{equation}

\noindent where the indices \, $\ell = e, \mu, \tau$ \, correspond to
the states which are not the mass eigenstates and are included
into the same representations as the charged leptons $\, \ell$

\begin{equation}
\nu_\ell \, = \, {\cal K}_{\ell i} \nu_i , \quad u_{\ell} \, = \,
{\cal U}_{\ell p} u_p , \quad d_{\ell} \, = \, {\cal D}_{\ell n}
d_n . \label{eq:nu1}
\end{equation}

\noindent Here \, $\nu_i, u_p, $ and $d_n$ \, are the mass eigenstates

\begin{eqnarray}
\nu_i \, = \, (\nu_1, \, \nu_2, \, \nu_3), \quad u_p \, = \,
(u_1, \, u_2, \, u_3), \, = \, (u, \, c, \, t), \nonumber \\
d_n \, = \, (d_1, \, d_2, \, d_3), \, = \, (d, \, s, \, b),
\label{eq:nu2}
\end{eqnarray}

\noindent and ${\cal K}_{\ell i} , \, {\cal U}_{\ell p}$, and
${\cal D}_{\ell n}$ \, are the unitary mixing matrices.

The well-known Lagrangian of the interaction of the charged weak currents
with the $W$ bosons in our notations has the form

\begin{eqnarray}
{\cal L}_W & = & \frac{g}{2 \sqrt 2} \big [
\big ( \bar \nu_{\ell} O_{\alpha} \ell \big ) +
\big ( \bar u_{\ell} O_{\alpha} d_{\ell} \big ) \big ] W^*_{\alpha} +
h.c. \, = \nonumber \\
& = & \frac{g}{2 \sqrt 2} \big [ {\cal K}^*_{\ell i}
\big ( \bar \nu_i O_{\alpha} \ell \big ) + {\cal U}^*_{\ell p} \;
{\cal D}_{\ell n} \;
\big ( \bar u_p O_{\alpha} d_n \big ) \big ] W^*_{\alpha} + h.c.,
\label{eq:Lw}
\end{eqnarray}

\noindent where $g$ is the $SU(2)_L$ group constant, and \,$O_{\alpha} \,
= \, {\gamma}_{\alpha}(1-{\gamma}_5)$. The standard Cabibbo--Kobayashi--
Maskawa matrix is thus seen to be $V \, = \, {\cal U}^+ \cal D$.
This is as far as we know about $\cal U$ and $\cal D$ matrices.
The $\cal K$ matrix description of the mixing in the lepton sector has been
the object of intensive experimental investigations in recent years.

Subsequent to the spontaneous $SU(4)_V$ symmetry breaking up to $SU(3)_c$
on the $M_X$ scale six massive vector bosons are separated from the 15-plet
of the gauge fields to generate three charged and colored leptoquarks.
Their interaction with the fermions~(\ref{eq:nu2}) has the form

\begin{equation}
{\cal L}_X \, = \, \frac{g_S(M_X)}{\sqrt 2} \big [
{\cal D}_{\ell n}
\big ( \bar \ell \gamma_{\alpha} d^c_n \big ) +
\big ( {\cal K^+ \; \cal U} \big )_{i p}
\big ( \bar{\nu_i} \gamma_{\alpha} u^c_p \big ) \big ] X^c_{\alpha} +
h.c. \,,
\label{eq:Lx}
\end{equation}

\noindent where the color index $c$ is written once again. The constant
\, $g_S(M_X)$ \, can be expressed in terms of the strong coupling constant
\, $\alpha_S$ \, at the leptoquark mass scale
$M_X, \quad g_S^2(M_X)/4 \pi = \alpha_S(M_X)$.

If the momentum transferred is \, $q \ll M_X$, \, then the Lagrangian
{}~(\ref{eq:Lx}) in second order leads to the effective four-fermion
vector-vector interaction of quarks and leptons. By using the Fiertz
transformation, lepton-current-to-quark-current terms of the scalar,
pseudoscalar, vector and axial-vector types may be separated in the
effective Lagrangian. Let us note that the construction of the effective
lepton-quark interaction Lagrangian requires taking account of the QCD
corrections estimated by known techniques~\cite{Vai,Vys}.
In our case the leading log approximation $ln(M_X/\mu) \gg 1$ with
$\mu \sim 1~GeV$ to be the typical hadronic scale is quite applicable.
Then the QCD correction amounts to the appearance of the magnifying factor
$Q(\mu)$ at the scalar and pseudoscalar terms

\begin{equation}
Q(\mu) \, = \, \left ( \frac{\alpha_S(\mu)}
{\alpha_S(M_X)} \right )^{4/\bar b} \, .
\label{eq:Qmu}
\end{equation}

\noindent Here $\alpha_S(\mu)$ is the effective strong coupling constant
at the hadron mass $ \, \mu \,$ scale,
$\; \bar b \, = \, 11 \, - \, \frac {2}{3} \bar n_f, \; \bar n_f $
is the averaged number of the quark
flavors at the scales $\mu^2 \le q^2 \le M_X^2$. If the condition
$M_X^2 \gg m_t^2$ is valid, then we have $\, \bar n_f \, \simeq \, 6$,
and $\bar b \, \simeq \, 7$.

It is interesting to investigate the contribution of the leptoquark
interaction~(\ref{eq:Lx})
to the low-energy processes in order to establish
the bounds on the model parameters from existing experimental limits.
As the analysis shows, the tightest restrictions on the leptoquark mass
$M_X$ and the mixing matrix elements $\cal D$
can be obtained from experimental data on rare $\pi$ and $K$ decays and
$\mu^- \rightarrow e^-$ conversion in nuclei. They are represented in
table 1. The amplitudes of these processes can be found in our
paper~\cite{Kuz}. Compared to ref.~\citenum{Kuz} we obtain here the improved
bounds based on the recent experimental data from TRIUMF, PSI, and BNL.

One can see from table 1 that the restrictions on the model parameters
contain the elements of the
unknown unitary mixing matrices $\cal D$ and $\cal U$, which are connected
by the condition ${\cal U}^+ {\cal D} = V$ only.
Thus the possibility is not excluded, in principle, that the bounds obtained
did not restrict $\, M_X \,$ at all, e.g. if the elements ${\cal D}_{e d}$
and ${\cal D}_{\mu d}$ were rather small. It would correspond to the
connection of the $\tau$ lepton largely with the $d$ quark in the ${\cal D}$
matrix, and the electron and the muon with the $s$ and $b$ quarks.
In general, it is not contradictory to anything even if it appears to be
unusual.

{\scriptsize
\begin{table}[ht]
\caption{The bounds on the leptoquark mass and mixing matrix
elements from the ex\-pe\-ri\-men\-tal limits on the branching ratios
of various processes.}

\vspace{4mm}

\begin{center}
\begin{tabular}{cccc}\hline \\
No. & Experimental limit & Ref. & Bound \\ \\ \hline \\
\bigskip
1 & $\frac{\mbox{\normalsize $\Gamma(\pi \rightarrow e \nu)$}}
{\mbox{\normalsize $\Gamma(\pi \rightarrow \mu \nu)$}}
= (1.2310 \pm 0.0037) \cdot 10^{-4}$ &
\citenum{Bri} &
$\frac{\mbox{\normalsize $M_X$}}
{\mbox{\normalsize $|Re({\cal D}_{e d} {\cal U}^*_{e u}/V_{u d}) |^{1/2}$}} \,
> \, 210~TeV$ \\
\bigskip
2 & $Br(K^+ \rightarrow \pi^+ \mu^- e^+ ) < 7 \cdot 10^{-9}$ &
\citenum{Dia} &
$\frac{\mbox{\normalsize $M_X$}}
{\mbox{\normalsize $|{\cal D}_{e s} {\cal D}^*_{\mu d} |^{1/2}$}} \, > \,
50~TeV$ \\
\bigskip
3 & $Br(K^+ \rightarrow \pi^+ \mu^+ e^- ) < 2.1 \cdot 10^{-10}$ &
\citenum{Lee} &
$\frac{\mbox{\normalsize $M_X$}}
{\mbox{\normalsize $|{\cal D}_{e d} {\cal D}^*_{\mu s} |^{1/2}$}} \, > \,
120~TeV$ \\
\bigskip
4 & $Br(K^0_L \rightarrow \mu^+ \mu^-) = (7.3 \pm 0.4) \cdot 10^{-9}$ &
\citenum{PDG} &
$\frac{\mbox{\normalsize $M_X$}}
{\mbox{\normalsize $|Re({\cal D}_{\mu d} {\cal D}_{\mu s}^*)|^{1/2}$}} \,
> \, 500 \div 600~TeV$ \\
\bigskip
5 & $Br(K^0_L \rightarrow \mu e ) < 3.3 \cdot 10^{-11}$ &
\citenum{Ari} &
$\frac{\mbox{\normalsize $M_X$}}
{\mbox{\normalsize $|{\cal D}_{e d} {\cal D}^*_{\mu s} \; + \;
{\cal D}_{e s} {\cal D}^*_{\mu d} |^{1/2}$}} \, > \, 1200~TeV$ \\
\bigskip
6 & $Br(K^0_L \rightarrow e^+e^- ) < 5.3 \cdot 10^{-11}$ &
\citenum{Ari} &
$\frac{\mbox{\normalsize $M_X$}}
{\mbox{\normalsize $|Re({\cal D}_{e d} {\cal D}^*_{e s})|^{1/2}$}} \, > \,
1400~TeV$ \\
\bigskip
7 & $\frac{\mbox{\normalsize $\Gamma(\mu^-Ti \rightarrow e^-Ti)$}}
{\mbox{\normalsize $\Gamma(\mu^-Ti \rightarrow capture)$}}
< 4.6 \cdot 10^{-12}$ &
\citenum{Ah} &
$\frac{\mbox{\normalsize $M_X$}}
{\mbox{\normalsize $|{\cal D}_{e d} {\cal D}^*_{\mu d} |^{1/2}$}} \, > \,
670~TeV$ \\
\hline
\end{tabular}
\end{center}

\end{table}
}
In this case a leptoquark could give a more noticeable
contribution to the flavor-changing decays of the $\tau$ lepton and
$\eta, \eta', \Phi$ and
$B$ mesons. However, a relatively poor accuracy of these data doesn't
yet allow to restrict the parameters essentially.

We could find only one occasion when the mixing-independent lower bound
on the leptoquark mass arises, namely, from the decay
$\pi^0 \rightarrow \nu \bar \nu$. In the paper~\cite{Lam} the cosmological
estimation of the width of this decay was found

\medskip

$Br(\pi^0 \rightarrow \nu \bar \nu ) < 2.9 \cdot 10^{-13}$ .

\medskip

\noindent Within the Standard Model this value is proportional to $m^2_\nu$.
The process is also possible through the leptoquark mediation, without the
suppression by the smallness of neutrino mass.
On summation over all neutrino species the decay probability
is mixing-independent. As a result the bound on the leptoquark mass is

\begin{equation}
M_X > 18~TeV.
\label{eq:X4}
\end{equation}

\medskip
In conclusion, we have analysed in detail the experimental data
on rare $\pi$ and $K$ decays and $\mu e$ conversion and we
have found the restrictions on the vector leptoquark mass to contain the
elements of an unknown mixing matrix $\cal D$. The only mixing independent
bound (\ref{eq:X4}) arises from cosmological estimations.

In our opinion, possible experimental manifestations of the considered
minimal quark-lepton symmetry model would be an object of further
methodical studies. For example, the search for possible leptoquark evidence
in the $p \bar p$ collider high-energy experiments via the reactions
$d \bar d \rightarrow e^+ \mu^-, \; e^- \mu^+$ could be of interest.
On the other hand,
further searches of flavor-changing decays of $\tau$ lepton and $\eta, \eta',
\Phi$ and $B$ mesons are desirable.

\vglue 0.6cm
{\elevenbf \noindent Acknowledgements \hfil}
\vglue 0.4cm

The authors are grateful to
L.B.~Okun, V.A.~Rubakov, K.A.~Ter-Martirosian, and A.D.~Smir\-nov
for fruitful discussions.

The research described in this publication was made possible in part by
Grant N RO3000 from the International Science Foundation.

\vglue 0.6cm
{\elevenbf\noindent References \hfil}
\vglue 0.4cm

\end{document}